\documentclass[conference]{IEEEtran}
\IEEEoverridecommandlockouts
\usepackage{cite}
\usepackage{amsmath,amssymb,amsfonts}
\usepackage{algorithmic}
\usepackage{graphicx}
\usepackage[binary-units]{siunitx}
\usepackage{textcomp}
\usepackage{xcolor}
\usepackage{listings}
\usepackage{booktabs}

\usepackage[T1]{fontenc}
\usepackage[utf8]{inputenc}

\usepackage[hidelinks]{hyperref} 
\usepackage{orcidlink}

\definecolor{codegreen}{RGB}{133, 153, 0}
\definecolor{codeblue}{RGB}{38, 139, 210}
\definecolor{codegray}{rgb}{0.5,0.5,0.5}
\definecolor{codeblack}{RGB}{7, 54, 66}
\definecolor{codepurple}{RGB}{211, 54, 130}
\definecolor{backcolour}{RGB}{253, 246, 227}

\lstdefinestyle{mystyle}{
    backgroundcolor=\color{backcolour},   
    commentstyle=\color{codegreen},
    keywordstyle=\color{codeblue},
    numberstyle=\tiny\color{codegray},
    stringstyle=\color{codepurple},
    basicstyle=\linespread{1.1}\ttfamily\footnotesize,
    emph={__global__, __shared__, __device__, __host__},
    emphstyle={\color{codegreen}},
    breakatwhitespace=false,         
    breaklines=true,                 
    captionpos=b,                    
    keepspaces=true,                 
    numbers=left,                    
    numbersep=5pt,
    xleftmargin=2em,
    framexleftmargin=2em,
    showspaces=false,                
    showstringspaces=false,
    showtabs=false,                  
    tabsize=2
}

\def\BibTeX{{\rm B\kern-.05em{\sc i\kern-.025em b}\kern-.08em
    T\kern-.1667em\lower.7ex\hbox{E}\kern-.125emX}}
\begin{document}

\title{Accelerating X-Ray Tracing for Exascale Systems using Kokkos}

\author{\IEEEauthorblockN{Felix Wittwer\,\orcidlink{0000-0002-5050-1669}}
\IEEEauthorblockA{\textit{National Energy Research Scientific Computing Center,} \\
\textit{Lawrence Berkeley National Laboratory,}\\
1 Cyclotron Road, Berkeley, CA 94720, USA \\
fwittwer@lbl.gov}
\and
\IEEEauthorblockN{Nicholas K. Sauter\,\orcidlink{0000-0003-2786-6552}}
\IEEEauthorblockA{\textit{Molecular Biophysics and Integrated Bioimaging Division,} \\
\textit{Lawrence Berkeley National Laboratory,}\\
1 Cyclotron Road, Berkeley, CA 94720, USA \\
nksauter@lbl.gov}
\and
\IEEEauthorblockN{Derek Mendez}
\IEEEauthorblockA{\textit{Molecular Biophysics and Integrated Bioimaging Division,} \\
\textit{Lawrence Berkeley National Laboratory,}\\
1 Cyclotron Road, Berkeley, CA 94720, USA \\
dermen@lbl.gov}
\and
\IEEEauthorblockN{Billy K. Poon\,\orcidlink{0000-0001-9633-6067}}
\IEEEauthorblockA{\textit{Molecular Biophysics and Integrated Bioimaging Division,} \\
\textit{Lawrence Berkeley National Laboratory,}\\
1 Cyclotron Road, Berkeley, CA 94720, USA \\
bkpoon@lbl.gov}
\and
\IEEEauthorblockN{Aaron S. Brewster\,\orcidlink{0000-0002-0908-7822}}
\IEEEauthorblockA{\textit{Molecular Biophysics and Integrated Bioimaging Division,} \\
\textit{Lawrence Berkeley National Laboratory,}\\
1 Cyclotron Road, Berkeley, CA 94720, USA \\
asbrewster@lbl.gov}
\and
\IEEEauthorblockN{James M. Holton\,\orcidlink{0000-0002-0596-0137}}
\IEEEauthorblockA{\textit{Molecular Biophysics and Integrated Bioimaging Division,} \\
\textit{Lawrence Berkeley National Laboratory,}\\
1 Cyclotron Road, Berkeley, CA 94720, USA \\
jmholton@lbl.gov}
\and
\IEEEauthorblockN{Michael E. Wall\,\orcidlink{0000-0003-1000-688X}}
\IEEEauthorblockA{\textit{Computer, Computational, and Statistical Sciences Division,} \\
\textit{Los Alamos National Laboratory,}\\
Mail Stop B256, Los Alamos, NM 87545, USA \\
mewall@lanl.gov}
\and
\IEEEauthorblockN{William E. Hart\,\orcidlink{0000-0002-6849-2780}}
\IEEEauthorblockA{\hspace{2cm}\textit{Sandia National Laboratories,}\hspace{2cm} \\
Albuquerque, NM 87185, USA \\
wehart@sandia.gov}
\and
\IEEEauthorblockN{Deborah J. Bard\,\orcidlink{0000-0002-5162-5153}}
\IEEEauthorblockA{\textit{National Energy Research Scientific Computing Center,} \\
\textit{Lawrence Berkeley National Laboratory,}\\
1 Cyclotron Road, Berkeley, CA 94720, USA \\
djbard@lbl.gov}
\and
\IEEEauthorblockN{Johannes P. Blaschke\,\orcidlink{0000-0002-6024-3990}}
\IEEEauthorblockA{\textit{National Energy Research Scientific Computing Center,} \\
\textit{Lawrence Berkeley National Laboratory,}\\
1 Cyclotron Road, Berkeley, CA 94720, USA \\
jpblaschke@lbl.gov}
}

\maketitle

\begin{abstract}
The upcoming exascale computing systems Frontier and Aurora will draw much of their computing power from GPU accelerators. 
The hardware for these systems will be provided by AMD and Intel, respectively, each supporting their own GPU programming model.
The challenge for applications that harness one of these exascale systems will be to avoid lock-in and to preserve performance portability.

We report here on our results of using Kokkos to accelerate a real-world application on NERSC's Perlmutter Phase 1 (using NVIDIA A100 accelerators) and the testbed system for OLCF's Frontier (using AMD MI250X).
By porting to Kokkos, we were able to successfully run the same X-ray tracing code on both systems and achieved speed-ups between 13\,\% and 66\,\% compared to the original CUDA code.
These results are a highly encouraging demonstration of using Kokkos to accelerate production science code.
\end{abstract}


\section{introduction}
The upcoming HPC systems Frontier and Aurora will be the first exascale computing systems.
Both are capable of performing more than \num{e18} floating point operations per second.
The majority of this computing power comes from the GPU accelerators of these systems.
Due to their massive parallelism, GPUs are well suited for repetitive tasks.

Using GPUs requires vendor specific programming models, such as CUDA for NVIDIA or HIP for AMD.
This makes portability between different systems challenging.
As an alternative, programming models such as OpenMP offloading or Kokkos provide an abstraction layer between the source code and the GPU hardware \cite{trottKokkosProgrammingModel2022}.
With these abstraction layers, the same code can be compiled for different architectures, thus combining portability with high performance.

\section{scientific background}
X-ray crystallography is an indispensable tool to study the structure of molecules, with applications ranging from material science \cite{schrieber2022} to understanding the function of proteins \cite{keable2021}.
In crystallography, small crystals of an unknown sample (e.g. a protein whose molecular structure is to be determined) are placed in an x-ray beam.
By measuring how the x-rays are scattered, the position of each atom in the sample can be determined.
A major application of x-ray crystallography is determining molecular structures of proteins.
Knowing the protein structure is crucial to understand how a protein interacts.
From this understanding, therapeutic drugs can be designed to, e.g., treat infections by blocking specific interactions.

\begin{figure}
    \centering
    \includegraphics{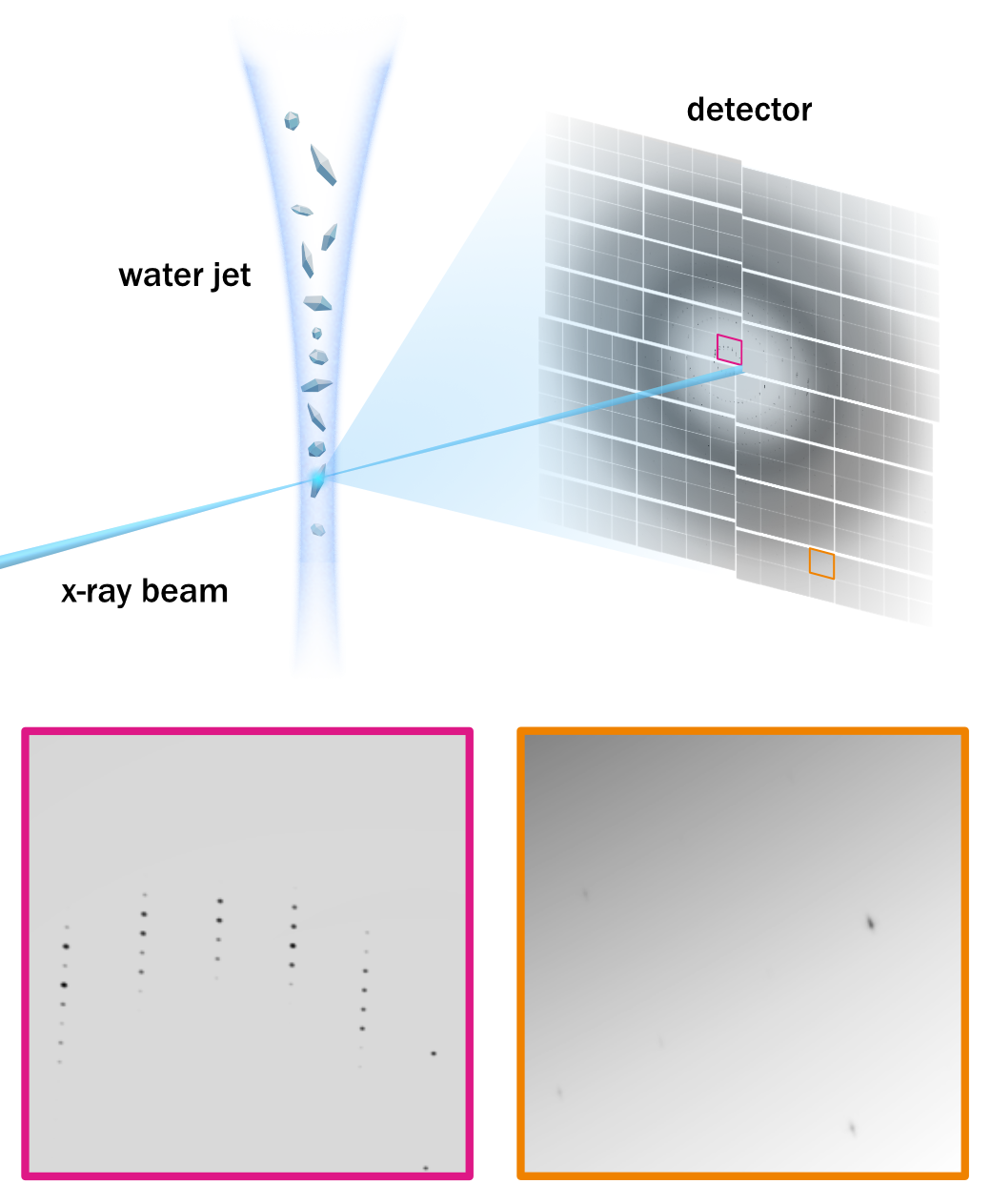}
    \caption{
    \emph{Top:} Experimental setup for SFX. 
    The protein crystals are delivered to the x-ray beam with a water jet.
    For each x-ray pulse, the imaging detector records the scattered x-rays.
    \emph{Bottom:} Two detector close-ups. Left: Small scattering angles close to the beam axis. Right: Wide angle scattering at the edge of the detector.}
    \label{fig:sfx_setup}
\end{figure}

Because the molecules inside the crystals are arranged in a long-range repeating order, the crystals scatter x-rays only into certain directions -- ie. scattering only occurs at those angles, which allow for positive interference from the x-rays scattered by different atoms.
X-rays scattered into these discrete directions form Bragg spots on the imaging detector (the bottom panels in Fig.~\ref{fig:sfx_setup} illustrate one possible Bragg spot pattern).
Mathematically speaking, a molecular structure is represented using a list of real numbers called structure factors, where each Bragg spot is associated with one such structure factor.
Reconstructing the protein structure is therefore a two step process: first, the structure factors are determined from the Bragg spots.
Then, the structure factors are used to reconstruct the protein structure.
If more structure factors are available for the reconstruction, the atomic positions can be determined more accurately.
Therefore, increasing the signal of weak Bragg spots requires larger crystals or a more intense x-ray beam.
However, many protein crystals are challenging to crystallize and grow only to microscopic size. Scientists therefore opt for increasing beam intensities.
However, as proteins are sensitive to radiation damage, they are quickly destroyed by high intensity x-ray beams.

Serial femtosecond crystallography (SFX) avoids the problem of radiation damage by using ultra-short x-ray pulses from x-ray free-electron lasers (XFELs), called ``shots''.
In essence, the pulses are so short that the Bragg spots can be measured before beam damage has hat time to degrade the sample.
Each x-ray pulse lasts only a few ten femtoseconds, fast enough to freeze all atomic motion and capture a snapshot of the crystal before any damage becomes visible.
Still, this process will destroy the sample, so fresh crystals must be constantly fed in before each x-ray pulse, see Fig.~\ref{fig:sfx_setup}.
Measuring a complete dataset requires thousands of crystals, each producing only a single scattering image.

Most crystallography methods determine the structure factors by simply integrating the number of photons in each Bragg spot.
Apart from the structure factors, the intensity of a Bragg spot is also influenced by the orientation of the crystal, how the crystal is composed from smaller crystals and how the energy spectrum of the photons fluctuates from shot to shot.
By averaging multiple images, each under different conditions, these influences can be averaged out.
For SFX, this requires tens of thousands of scattering images.

Instead of simply integrating each Bragg spot, x-ray tracing aims to simulate the photon intensity of every detector pixel by creating a physical model of each crystal \cite{mendezIntegrationModelingEvery2020}. 
Modeling each crystal is an iterative process, in which the current parameter estimates are refined until the simulated scattering image matches the measured one.
By recovering the unknown parameters from each crystal, x-ray tracing can accurately determine the structure factors from an order of magnitude fewer scattering images.
As measurement time at XFELs is scarce, this allows scientists to study more samples or more experimental conditions during the same experiment.

However, while x-ray tracing may require fewer experimental resources, the computational effort to accurately model each pixel grows considerably in comparison to conventional methods.
X-ray tracing therefore faces the challenge of performing a more complex data analysis in less time.
At the same time, quick feedback on the quality of the collected data is critical for XFEL experiments \cite{blaschkeRealtimeXFELData2021}.
Providing these experiments with quick-turnaround analysis of the terabytes of data requires high performance computing (HPC) at the Exascale.

\section{gpu acceleration with kokkos}
To simulate scattering images from a physical model, we have developed the program \emph{nanoBragg} \cite{sauterSpatialResolutionMetalloprotein2020}, which is part of the CCTBX software suite\cite{grosse2002,sauter2013}.
CCTBX consists of a high-level workflow written in Python and accelerated kernels written in C++ (interfacing with Python via \emph{Boost.Python}).
This approach has been proven to be highly successful in analyzing large data sets using HPC resources \cite{blaschkeRealtimeXFELData2021,giannakou2021}. 
By designing \emph{nanoBragg} as a kernel compatible with the CCTBX workflow, we see it as the first building block of a larger pixel-level XFEL data analysis workflow.

In x-ray tracing, each detector pixel can be simulated nearly independently of all other pixels, making x-ray tracing ideally suited to exploit the massive parallelism of modern graphic processing units (GPUs).
To this end, the original \emph{nanoBragg} code was previously ported to NVIDIA CUDA, resulting in a 500$\times$ speedup on an NVIDIA V100 over a multi-threaded CPU implementation on Intel Xeon Ivy Bridge.
\emph{nanoBragg} distributes tasks (batches of images to simulate) to multiple nodes and GPUs using MPI (using \emph{mpi4py}).
Each GPU then performs work independently.
Solving for an unknown molecular structure therefore follows a fork-join parallelism, where each GPU simulates a batch of parameters independently, followed by a MPI reduction\cite{mendezIntegrationModelingEvery2020}.
The \emph{nanoBragg} benchmark discussed here forgoes the final reduction step, instead saving the simulated images to the file system.

The upcoming exascale systems Frontier (OLCF) and Aurora (ALCF) are essential to obtain rapid feedback during XFEL experiments.
However, the GPUs for these systems will be provided by AMD, and Intel, respectively.
As both vendors supply their own alternative to CUDA, \emph{nanoBragg} must be adapted to each system.
To avoid code duplication and site-specific optimizations, we decided to use Kokkos \cite{trottKokkosProgrammingModel2022}.

\subsection{Kokkos}
Kokkos is a C++ programming model that allows a single code base to target different HPC platforms \cite{trottKokkosProgrammingModel2022}.
To achieve this, Kokkos implements abstract memory and execution spaces.
Only during the compilation are the abstract spaces and commands converted to CUDA, HIP, OpenMP, etc.
This way, the same Kokkos code can be used on systems with different GPUs or even systems with no GPU.

Each execution space is associated with a particular memory space.
While the execution space defines where the computation is done, the computation itself is defined by \emph{execution patterns}, which are the Kokkos equivalent to kernels in CUDA.
Kokkos implements three patterns:
\begin{itemize}
    \item \verb!parallel_for!: Corresponds to a for-loop where each iteration is executed independently, for example element-wise addition of two vectors.
    \item \verb!parallel_reduce!: A reduction that collects and combines the result of all iterations, for example finding the longest word in a list.
    \item \verb!parallel_scan!: A combination of the former two that uses multiple reductions, for example calculating a histogram of an image.
\end{itemize}

The typical use case for execution patterns is to work on large data arrays.
Kokkos provides its own data structure called \emph{View}.
A \emph{View} is a multi-dimensional array which can be transferred between different memory spaces and layouts.
The last one is important, e.g. in matrix multiplications, the best performance on a multi-core CPU might be achieved with a row-major layout and on a GPU with a column-major layout.
Kokkos automatically manages the conversion between layouts when a \emph{View} is transferred between different memory spaces.
This makes it easy to first initialize a \emph{View} on the CPU and then transfer it to GPU memory.

\subsection{Porting nanoBragg to Kokkos}
The task of porting \emph{nanoBragg} can be split into two parts: The first part is converting the CUDA kernels to Kokkos' execution patterns, the second part is replacing all CUDA arrays with Kokkos' \emph{Views}.
As there is no interaction between the individual detector pixels, all kernels can be replaced with the parallel\_for pattern.
Most of the computational work in \emph{nanoBragg} is done in three Kernels:
\begin{itemize}
    \item \emph{nanoBraggSpots}: The most complex kernel, containing about 350 lines of code. This kernel simulates the Bragg spots. It takes into account crystal orientation, mosaicity and photon energy.
    \item \emph{addBackground}: The second most complex kernel with about 120 lines of code. This kernel simulates the background scattering from the air and water that surround the crystal.
    \item \emph{addArray}: A simple kernel that adds the results from the other two kernels.
\end{itemize}

We illustrate the porting process by using a simplified version of the \emph{addArray} kernel:
\begin{lstlisting}[language=C++, caption=The \emph{addArray} kernel in CUDA]
__global__
void addArray(double* lhs,
              float* rhs, 
              int total_pixels) {
  int j = blockDim.x * blockIdx.x + threadIdx.x;
  if (j < total_pixels) {
    lhs[j] = lhs[j] + (double) rhs[j];
  }
}  
\end{lstlisting}

For the Kokkos version, we replaced the \emph{lhs} and \emph{rhs} arrays with \emph{Views} and the body of the kernel with a parallel\_for pattern.
As the memory management of \emph{Views} is done by Kokkos, we could remove the custom-written CUDA memory management.
The parallel\_for execution pattern takes three arguments: an individual name for debugging and profiling purposes, an execution policy and a functor that holds the computational work.
In simple cases like this, the execution policy is simply to iterate over all \emph{total\_pixels}.
The functor can be integrated into the pattern by using lambda expressions, this way the structure mirrors how kernels are defined:
\vspace{1cm}
\begin{lstlisting}[language=C++, caption=The \emph{addArray} kernel in Kokkos]
void addArray(Kokkos::View<double*> lhs,
              Kokkos::View<float*> rhs,
              int total_pixels) {
  Kokkos::parallel_for(
    "addArray", total_pixels, 
    KOKKOS_LAMBDA (const int& j) {
      lhs(j) = lhs(j) + (double) rhs(j);
    }
  );
}
\end{lstlisting}

The other kernels where converted much in the same way, with a few challenges.
One of these is an edge case when using lambda expressions for device code in member functions of a class.
Suppose \emph{init()} is a member function that initializes the view \verb!data! with a certain value:
\begin{lstlisting}[language=C++, caption=Lambda capture]
void Container::init() {
  Kokkos::parallel_for("init", size, 
    KOKKOS_LAMBDA (const int& j) {
      data(j) = m_value;
    }
  );
}
\end{lstlisting}
Since they are class members, the compiler replaces \verb!data! and \verb!m_value! during the compilation with \verb!this->data! and \verb!this->m_value!.
The problem arises from \verb!this! being a pointer in CPU memory.
If CUDA is used, this creates an illegal memory access when the GPU tries to access the pointer.

There are multiple ways around this issue: One way is to avoid lambda expressions in this situation and instead explicitly define the functors.
Another option is to copy the specific member variables into local variables before the execution pattern is called.
And finally, in C++17 lambda expressions have been extended so they can explicitly capture the object, not the pointer. 
Unfortunately, the CCTBX code is currently not compatible with C++17.
Therefore, we choose to use local variables in these situations and otherwise minimize the use of kernels in member functions.



\section{Performance and Portability}
\label{sec:performance}

\begin{figure}
	\centering
	\includegraphics[width=\linewidth]{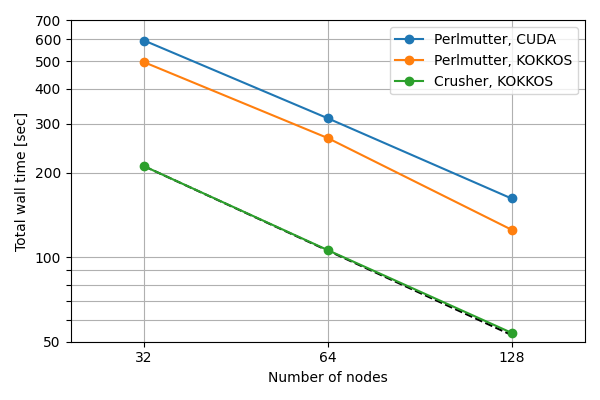}
	\caption{
		Strong scaling of \emph{nanoBragg} for the simulation of \num{100000} scattering images.
		The dashed line shows ideal scaling where doubling the number of nodes halves the simulation time.
		On Perlmutter Phase 1, Kokkos is used with the CUDA backend.
		On Crusher, Kokkos is used with the HIP backend.
	}
	\label{fig:strong_scaling}
\end{figure}
The goal of porting the \emph{nanoBragg} code to Kokkos is to achieve portability between the upcoming Frontier and Aurora systems as well as Perlmutter.
We test the portability on Perlmutter Phase 1 and Crusher, the latest Frontier testbed system.
These systems are not production resources, the performance numbers given here were not always reproducible and will be different from the final systems.

Increasing the portability of a code can potentially reduce the performance.
We therefore also assess on Perlmutter the performance of the Kokkos port in comparison to the original CUDA version.
On Perlmutter, we use the CUDA backend of Kokkos and on Crusher accordingly the HIP backend.

Each node of Perlmutter Phase 1 is equipped with an AMD EPYC 7763 CPU and four NVIDIA A100 GPUs.
On Crusher each node is equipped with an AMD EPYC 7A53 CPU and four AMD MI250X GPUs.
Each MI250X GPU is equipped with two graphic compute dies (GCDs) totalling eight GCDs per node.
These eight GCDs function effectively like eight separate GPUs per node.

As a benchmark, we used a test scenario that simulates \num{100000} scattering patterns.
Due to performance fluctuations of the file system (a consequence of Perlmutter and Crusher not being production systems), we did not include the saving of the simulated images into the benchmark.
The tests were performed using 32, 64, and 128 nodes, all results are plotted in Fig.~\ref{fig:strong_scaling}.
Every test case shows almost ideal scaling behavior, highlighting how x-ray tracing benefits from parallel computing.
On Perlmutter, the Kokkos version is about \SI{13}{\percent} faster than the original CUDA version, indicating no loss in performance due to the port to Kokkos.
Concerning portability, Kokkos allows \emph{nanoBragg} to exploit the hardware of Crusher for a more than \SI{60}{\percent} faster performance per node, compared to Perlmutter Phase 1.

\begin{figure}
    \centering
	\includegraphics[width=\linewidth]{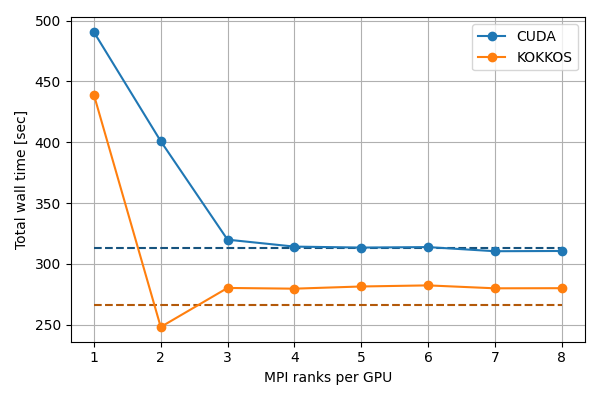}
	\caption{
		Time to simulate 100 thousand diffraction patterns using 64 Perlmutter nodes. In contrast to Fig.~\ref{fig:strong_scaling}, this benchmark includes the time to save each image to the file system. For comparison, the previous times without file saving are indicated with dashed lines. As the filesystem performance on Perlmutter can vary (due to it not being a production system) at the time of writing, this graph shows the best performance over several repeated runs.
	}
	\label{fig:multi_tenancy}
\end{figure}

Next we re-enable file I/O. This introduces workflow latency and dependence on a shared resource (the file system).
We observe that the strong scaling in shown in Fig.~\ref{fig:strong_scaling} is preserved~--~albeit shifted up due to the increased time spent on I/O.
CCTBX seeks to hide workflow latency by assigning more than one MPI rank to each GPU.
We observe that this is a largely successful strategy (as total runtime increases with increasing number of MPI ranks per GPU).
Fig.~\ref{fig:multi_tenancy} shows the total runtime to trace 100 thousand images using 64 Perlmutter nodes.
We see that increasing the number of ranks per GPU increases the total throughput.
Interestingly, the CUDA implementation achieves the highest throughput when many (approx. 7 or greater) ranks are sharing the same GPU, whereas the Kokkos implementation achieves the maximum throughput when two MPI ranks share the same GPU.
Furthermore, comparing Fig.~\ref{fig:strong_scaling} and Fig.~\ref{fig:multi_tenancy}, we see that the increased latency due to I/O can be hidden by sharing GPUs accross ranks.

\section{Kernel profiling}
\begin{table}
    \centering
    \caption{Kernel run-times.}
    \begin{tabular}{c c c c} 
        \toprule
        & nanoBraggSpots & addBackground & addArray \\
        \midrule
        CUDA & \SI{8.28}{\ms} & \SI{1.87}{\ms} & \SI{0.13}{\ms} \\
        Kokkos & \SI{6.98}{\ms} & \SI{1.76}{\ms} & \SI{0.12}{\ms} \\
        Speed-up & +\SI{+15.7}{\percent} & +\SI{5.9}{\percent} & +\SI{7.7}{\percent} \\
        \bottomrule
    \end{tabular}
    \label{tab:kernel_times}
\end{table}
To analyze how the Kokkos version performs better on Perlmutter than the original CUDA version, we used Nsight Systems and Nsight Compute to profile both versions.
For the three main kernels, the execution times on a NVIDIA A100 GPU are given in Table~\ref{tab:kernel_times}.
For all three kernels, Kokkos achieves a speed-up of more than five percent.
As the \emph{nanoBraggSpots} kernel accounts for most of the GPU computation time, the speed-up of more than \SI{15}{\percent} for this kernel has a significant impact on the overall run time.
On the other end of the complexity spectrum is the \emph{addArray} kernel.
While it is one of the smallest kernels and any improvements will only have a small influence in the grand scheme, achieving a speed-up of nearly eight percent even for short kernels is remarkable.

\begin{table}
    \centering
    \caption{nanoBraggSpots details.}
    \begin{tabular}{r c c} 
        \toprule
        & CUDA & Kokkos \\
        \midrule
        Run-time & \SI{8.28}{\ms} & \SI{6.98}{\ms} \\
        Compute Throughput & \SI{65.05}{\percent} & \SI{77.42}{\percent} \\
        Memory Throughput & \SI{21.05}{\percent} & \SI{21.35}{\percent} \\
        Registers & \num{130} & \num{116} \\
        Theoretical Occupancy & \SI{18.75}{\percent} & \SI{25}{\percent} \\
        Achieved Occupancy & \SI{16.8}{\percent} & \SI{24.74}{\percent} \\
        \bottomrule
    \end{tabular}
    \label{tab:kernel_details}
\end{table}
Using Nsight Compute we studied the critical \emph{nanoBragg\-Spots} kernel in detail.
The results are summarized in Table~\ref{tab:kernel_details}.
In the table, the compute and memory throughput give an overview how much of the compute and memory resources of the GPU are utilized by the \emph{nanoBraggSpots} kernel.
The performance difference between CUDA and Kokkos is mostly a result of the higher compute throughput of the Kokkos kernel, the memory throughput is nearly identical.
The increased compute throughput in Kokkos is made possible by using fewer registers.
Each multiprocessor on an A100 GPU has a total of \num{65536} \SI{32}{\bit}-registers for a maximum of \num{2048} simultaneous threads (64 warps per multiprocessor times 32 threads per warp).
However, the GPU can only run at full occupancy if each thread uses less then $65535/2048=32$ registers.
With \num{130} registers, the GPU can run at most \num{12} warps out of the \num{64} available.
The Kokkos kernel uses \num{116} registers, which is below the critical threshold of \num{128} registers, and can therefore run a maximum of \num{16} warps, \SI{33}{\percent} more.

The situation for the other kernels is similar.
The \emph{addBackground} kernel goes from \num{96} to \num{80} registers and the \emph{addArray} kernel goes from \num{25} to \num{16} registers.
Kokkos uses consistently fewer registers than the original CUDA implementation, allowing a higher occupancy and throughput.

\section{Conclusion}
In this paper, we have reported on our work to port \emph{nanoBragg} to Kokkos.
Starting from a CUDA code base, most of the port was straightforward.
Arising difficulties from the port could be solved.
We have demonstrated that the performance of the code did not suffer from this port and has even increased by \SI{13}{\percent}.
We have also successfully demonstrated performance portability between Perlmutter Phase 1 (NERSC) and Crusher (OLCF).
We are currently in the process of testing the Aurora test system (ALCF).
Until now, we have not used advanced Kokkos features such as nested parallelism, which should further increase the performance.

As \emph{nanoBragg} is representative of many scientific codes, our reported findings indicate that Kokkos is a powerful tool to not only achieve performance portability in real-world applications, but also to accelerate existing codes.
Moreover, we are able to integrate Kokkos into existing scientific workflows without needing to port an entire code base. 
This allows scientists to make progress, even if staff time is limited.

\section{Acknowledgments}
N.K.S, J.P.B., F.W. and D.B. acknowledge support from the Exascale Computing Project (grant 17-SC20-SC), a collaborative effort of the Department of Energy (DOE) Office of Science and the National Nuclear Security Administration.
This research used resources of the National Energy Research Scientific Computing Center (NERSC), a U.S. Department of Energy Office of Science User Facility located at Lawrence Berkeley National Laboratory, operated under Contract No. DE-AC02-05CH11231. 
This research used resources of the Oak Ridge Leadership Computing Facility at the Oak Ridge National Laboratory, which is supported by the Office of Science of the U.S. Department of Energy under Contract No. DE-AC05-00OR22725.

\bibliographystyle{ieeetr}
\bibliography{bibliography}

\end{document}